# Natural Asset Beta

**Author:** Daniel Grainger (daniel.a.grainger@gmail.com), Townsville, Australia.

**Highlights**

- An endogenous discounting sustainability model is extended to macroeconomies
- An asset beta corporate finance notion is developed for natural resources
- The link to market exchange price for natural resources is made
- The approach shows promise in a test on real world data
- Monetary accounting for natural capital accounting is thus facilitated

**Abstract**

Natural capital accounting is important to efforts that attempt to measure the value of nature to decide on how best to trade-off natural resource productive use and conservation. Much work on measuring reciprocal physical stocks and flows of natural resources between nature and the economy has occurred. However, the current open problem of translating natural resource physical stocks and flows into monetary ones presents a barrier to estimating opportunity costs to inform resource allocation decisions. The research reported here extends a theoretical model on economic sustainability and derives a novel asset beta notion for nature. Like traditional firm asset beta notions, this underpins corporate finance notions for nature. Supply and demand curves estimating the exchange value of the natural resource are possible. A real-world data set is utilised for a preliminary proof of concept test. The results show promise, but future research is suggested investigating the sensitivity of the estimates.

**JEL classification** D62, O44, Q01, Q20, Q32

## 1 Introduction

Natural capital accounting has built significant momentum in global efforts to account for and properly manage natural resources (Comte et al., 2022). Much has been done to measure physical stocks and flows of natural resources, but converting these into monetary units is an open problem (Keith et al., 2021, Hein et al., 2020). The failure to assign monetary values to the stocks and flows of natural resources impairs trade-off decisions that inform the allocation of natural resources (Brandon et al., 2021). This is particularly problematic for trade-off decisions that attempt to ensure intergenerational equity sustainability, which underpins a great mass of the work, inter alia, Dasgupta (2021), Leemans and De Groot (2003) and Accounting and Accounting (2014). The requisite natural resource opportunity cost estimates are assumed as known in the related theoretical work e.g., Solow (1974), Hartwick (1977), Hartwick (1978), Stollery (1998). The present paper is novel in that it relaxes this assumption and extends the theoretical model of Grainger (2024) to develop the necessary supply and demand schedules that imply market equilibrium exchange prices for natural resources. By extending the corporate finance asset beta notion into a natural asset beta for nature, the present paper provides an approach that is consistent with the current industry practice of enterprise valuation.

Thinking on how best to manage the conservation and productive use trade-off of natural resources has a long history that saw the substantive beginning in the seminal work of Hotelling (1931). Subsequent research by Solow (1974) to incorporate a Rawlsian notion of intergenerational equity (i.e., that no generation is better off than another and that there is effectively constant utility across time) stimulated further related work by Asheim (2010), Cairns and Van Long (2006) and Mitra et al. (2013). While the work of Stollery (1998) and d'Autume and Schubert (2008) provided a means to model nature-based features as entering directly into the individual utility function, an arguably more tractable way of incorporating nature into economic models was formulated in Hartwick and Van Long (2020) by restricting the individual utility function to be one which was simply a function of consumption. Grainger (2024) extended this constant consumption sustainability theoretical model of Hartwick and Van Long (2020) into a more general setting containing notions of externalities and elasticities. It also provided a starting point to consider corporate finance notions of risk-return in its development of a rate of return based sustainability criterion that was a result of general equilibrium market exchange prices.

Rate of return notions were implied in the economic models incorporating endogenous discounting of Le Kama and Schubert (2007), Obstfeld (1990) and Epstein (1987), but it was Hartwick and Van Long (2020) that made the explicit link to rate of returns being a function of macroeconomic variables. This is important because it allows the rate of return to differ across economies as a function of those macroeconomic variables and given those returns are not static (as is the case in formative articles of Hartwick (1977) and Hartwick (1978)), provides the ability to consider their covariation and hence arrive at corporate finance asset beta notions such as in Markowitz (1991) and Sharpe (1964).

This paper starts out by extending the single Grainger (2024) economy, with its single endogenous discount factor, into one with multiple neighbouring economies possessing different endogenous discount factors. This allows the ability to consider the variation between rates of returns and hence arrive at a novel asset beta notion for natural resources; called the natural asset beta. The notion of traditional firm asset betas is generalized into natural asset betas. Importantly, the link between natural asset betas and market exchange prices for natural resources to ensure constant consumption sustainability is made. Finally, a real-world data set provides a means to test the theory and perform a preliminary proof of concept. The estimates of natural asset beta, rate of return for the natural resource and the associated market exchange price is also reported.

## 2 The basic economic model

The Grainger (2024) economy contains infinitely lived individuals indexed by $\theta \in [0,1]$ and assumes zero population growth. There exists a strictly increasing and concave heterogeneous utility function $u_\theta(c(t,\theta))$ for arbitrary individual (or firm) $\theta$; where $c(t,\theta)$ denotes the consumptions of that individual at time $t$. Capital stock and natural resource stocks are owned by an individual and are respectively denoted $k(t,\theta)$ and $x_j(t,\theta)$ for various resource stocks indexed by $j \in \mathbb{N}$. Resource flows due to an individual extracting from stock $x_j(t,\theta)$ is denoted $q_j(t,\theta)$ with the dynamics of this extraction described by $\dot{x}_j(t,\theta) = G_j(x_j(t,\theta)) - q_j(t,\theta)$ where $G_j(x_j(t,\theta))$ is the natural resource growth and $\dot{x}_j(t,\theta)$ denotes the first derivative with respect to time of $x_j(t,\theta)$. Note that if $G_j(x_j(t,\theta)) \not> 0$ the resource is not renewable. The price of resource $j$ at time $t$ is denoted $p_j(t)$ and the rate of return on the capital stock is denoted $r(t)$. The individual's income and capital accumulation are respectively $y(t,\theta) = r(t)k(t,\theta) + \sum_j p_j(t)q_j(t,\theta)$ and $\dot{k}(t,\theta) = y(t,\theta) - c(t,\theta)$.

At the aggregate level the economy's capital stock, resource stock $j$, resource flows, consumption and income are respectively $K(t) = \int_0^1 k(t,\theta)d\theta$, $X_j(t) = \int_0^1 x_j(t,\theta)d\theta$, $Q_j(t) = \int_0^1 q_j(t,\theta)d\theta$, $C(t) = \int_0^1 c(t,\theta)d\theta$ and $Y(t) = \int_0^1 y(t,\theta)d\theta$. The inverse demand function of a natural resource is assumed to be strictly monotonic with respect to extraction of resources with $\frac{\partial p_j(t)}{\partial Q_k(t)}$ well behaved for all $k$. In this model, each individual maximizes the present value of utility flows by choosing paths for their consumption $c(t,\theta)$ and extraction for each resource $q_j(t,\theta)$ i.e., $max \int_0^\infty u_\theta(c(t,\theta))\beta(t)d\theta$ where $\beta(t)$ is the endogenous discount factor of Hartwick and Van Long (2020); where making the constant consumption assumption and assuming no resource growth implies that $\frac{\dot{\beta}}{\beta} = -r$.

Grainger (2024) finds, inter alia, that the Rawls-Solow-Hartwick constant consumption sustainability ($\dot{c} = 0$ ensuring intergenerational equity $\dot{u}_\theta = 0$) path, for which the aggregate level of consumption is maximised, is one on which the marginal user cost of remaining reserves equals the marginal user cost of extracted resources invested in reproducible capital at the financial rate of return net of the resource growth rate (more formally $-\frac{\partial(\sum_k p_k q_k)}{\partial x_j} = (r - G'_j)\frac{\partial(\sum_k p_k q_k)}{\partial q_j}$ at the economic equilibrium).

The modification to this model is to consider L neighbouring economies (of the form specified above) that differ in their endogenous discount factor $\beta(t)$. This is indeed possible because the endogenous discount factor is a function of the macroeconomic variables $K$, $Q$, $X$ and $C$ that can differ across the economies; with the single economy as specified in Hartwick and Van Long (2020). Ignoring resource growth for the time being, the utility in making this modification is that each economy can have different discount rates $r = -\frac{\dot{\beta}}{\beta}$ associated with their future cashflows and therefore the covariation in returns allows corporate finance asset beta notions to naturally fall out. The implications of this can be simplified by considering two of the L neighbouring economies interacting through resource flows across their borders. Formally, economy N (visualise an economy representing nature) and economy F (visualise the standard reproducible economy) sees the same cashflow $c$ associated with a resource flow in a different way due to their different rates of returns $r^{(N)}$ and $r^{(F)}$ respectively. To aid in the exposition, let the resource flow for resource $j$ be denoted $\Delta x_j$ in economy N and $q_j$ in economy F. Then the marginal user cost for the resource will be $\pi^{(N)}{}_{\Delta x_j} = \frac{\partial(\sum_k p_k q_k)}{\partial \Delta x_j}$ and $\pi^{(F)}{}_{q_j} = \frac{\partial(\sum_k p_k q_k)}{\partial q_j}$ in N and F respectively. The marginal user costs are related to the cashflow $c$ such that $c = \overline{\pi^{(N)}{}_{\Delta x_j}} = r^{(N)}\pi^{(N)}{}_{\Delta x_j} = r^{(F)}\pi^{(F)}{}_q = \overline{\pi^{(F)}{}_{q_j}}$. This implies that $\frac{r^{(N)}}{r^{(F)}}\pi^{(N)}{}_{\Delta x_j} = \pi^{(F)}{}_q$ where $\frac{r^{(N)}}{r^{(F)}}$ is defined as the asset beta denoted $\beta_{NF}$; noting that herein the use of $\beta$ to denote both asset betas and the endogenous discount factor that establishes $r$ is clear in the context. The economic implication is that the asset beta acts to translate the marginal opportunity cost when a unit of resource is used in each economy such that $\beta_{NF}\pi^{(N)}{}_{\Delta x_j} = \pi^{(F)}{}_q$. Notice that $\beta_{NF} < 1$ implies $r^{(N)} < r^{(F)}$ (i.e., future cashflows matter relatively more in N than in F) and $\pi^{(F)}{}_q < \pi^{(N)}{}_{\Delta x_j}$ (i.e., use of the resource results in forgoing more opportunities for a business (firm) in N than one in F). A more formal consideration of this, which includes resource growth, will be detailed below.

This provides a useful starting point to make connections to foundational notions of corporate finance beyond asset betas, namely the subsequent calculation of returns and market exchange price. Importantly, it suggests a way to develop natural asset betas that underpin the rate of return and market exchange price estimates for natural resources that are framed in terms of "nature" effectively acting as a business that is maintained by and benefits humans i.e., to align with the framing of natural capital accounting.

## 3 Connections to corporate finance for natural asset betas

A slight change in nomenclature is now performed to more easily view corporate finance elements of the Grainger (2024) relationship $-\frac{\partial(\sum_k p_k q_k)}{\partial x_j} = (r - G'_j)\frac{\partial(\sum_k p_k q_k)}{\partial q_j}$.

In so doing, on the right-hand side of the relationship, the rate of return associated with investing in a firm denoted $r_q = r - G'_j$ can be seen to be multiplied by the price of the resource flows out of the natural stock and into the firm denoted $\pi_q = \frac{\partial(\sum_k p_k q_k)}{\partial q_j}$ with user cost $\pi = \sum_k p_k q_k$ at time $t$. Let the present value of future extracted resources at $t = 0$ be $V = \int_0^\infty \pi \beta \, dt$ then the marginal (present) value of the remaining stock to the firm on the optimal path is $-V_x = \int_0^\infty -\pi_x \beta \, dt = \int_0^\infty \dot{\pi}_q \beta \, dt$ where $\dot{\pi}_q = r_q \pi_q$. Notice that integrand $\dot{\pi}_q$ can be interpreted as the future marginal (or per unit) cashflow generated by investment in the firm. Thus, $r_q \pi_q = (r - G'_j)\frac{\partial(\sum_k p_k q_k)}{\partial q_j}$ are marginal cashflows from investment into the firm of a unit of resource of value $\pi_q$ with rate of return $r_q$. Notice also that cashflow $-\frac{\partial(\sum_k p_k q_k)}{\partial x_j}$ can likewise be seen to result from an investment of a unit of resource into (N)atural stock of marginal value $\pi^{(N)}{}_{\Delta x}$ and rate of return $r_x$ i.e., $\pi^{(N)}{}_{\Delta x} r_x = -\frac{\partial(\sum_k p_k q_k)}{\partial x_j}$. In summary, the economic equilibrium relationship $-\frac{\partial(\sum_k p_k q_k)}{\partial x_j} = (r - G'_j)\frac{\partial(\sum_k p_k q_k)}{\partial q_j}$ becomes $r_x \pi^{(N)}{}_{\Delta x} = r_q \pi_q$; representing reciprocal cashflows between nature and the firm. This reasoning lends itself to the corporate finance notion of asset betas discussed in the following.

For a zero-valued risk-free rate of return, the firm's asset beta can be denoted $\beta_{qm} = \frac{E[r_q r_m]}{E[r_m r_m]}$; where $r_m$ denotes the market portfolio rate of return. This can be usefully simplified either by using a Taylor expansion for the expectation $E[r_q r_m]$ and $E[r_m r_m]$ about $r_q r_m = 0$ and $r_m r_m = 0$ respectively, to arrive at $\beta_{qm} = \frac{r_q r_m}{r_m r_m} = \frac{r_q}{r_m}$, or simply observing that the regression for $r_q = r_m \beta_{qm}$ provides a similar estimate $\beta_{qm} = [r'_m r_m]^{-1} r'_m r_q$. Similarly, $\beta_{xm} = \frac{r_x}{r_m}$. Notice that $\beta_{qm}$ is the standard notion of asset beta for the firm that informs the investor of the risk-return relationship for the operation of the firm as it extracts resources from nature. Notice also that $\beta_{xm}$ is the logical extension of the asset beta idea when considering nature as if it operated as a firm or investment (a central principle that underpins nature capital accounting). That is the natural asset beta $\beta_{xm}$ represents the risk-return relationship for investments in nature.

# 4 Connections to corporate finance valuation

It is informative to consider the covariation of the natural resource returns $r_x$ with respect to firm returns $r_q$ by defining $\beta_{xq} = \frac{\beta_{xm}}{\beta_{qm}} = \frac{\frac{r_x}{r_m}}{\frac{r_q}{r_m}} = \frac{r_x}{r_q}$. This has utility in the economic interpretation of the terms in the relationship $r_x \pi^{(N)}{}_{\Delta x} = r_q \pi_q$. Specifically, $r_q \pi_q$ on the right-hand side is the firm generated cashflows that can be seen to supply and compensate the cashflow demand and payment $r_x \pi^{(N)}{}_{\Delta x}$ to nature for a unit of its resources on the left-hand side of the relationship. The relationship may be rewritten as $r_q \left(\beta_{xq} \pi^{(N)}{}_{\Delta x}\right) = r_q \pi_q$ to more clearly show the risk adjusted marginal value (price) for nature $\beta_{xq} \pi^{(N)}{}_{\Delta x}$ and the firm $\pi_q$ are equal i.e., $\beta_{xq} \pi^{(N)}{}_{\Delta x} = \pi_q$. That is, $\beta_{xq}$ is utilised to adjust for risk differences in the nature and firm investment settings. Extending the thinking further, allow $\beta_{xq}, r_q$ and $r_x$ to be the usual corporate finance notions of constant valued asset beta and asset rate of returns. Thus, $r_x = \beta_{xq} r_q$ may be rewritten as $\overline{\left(\dot{\ln(\pi_{\Delta x})}\right)} = \beta_{xq} \overline{\left(\dot{\ln(\pi_q)}\right)}$ and then integrated to arrive at an equilibrium condition $\ln(\pi_{\Delta x}) = \beta_{xq} \ln(\pi_q) + d$ that results from the need to balance payments to nature $\pi_x$ for resource flows from nature to the firm $q$ (e.g., use of natural resources for production) with payments to the firm for resource flows from the firm to nature $\Delta x$ (e.g., arising from conservation of nature efforts). This can be represented by a system of simultaneous equations using variables $\widehat{\pi_q}, \widehat{\pi_{\Delta x}}, \hat{q}, \widehat{\Delta x}$ in first-order relationships $\ln(\widehat{\pi_{\Delta x}}) = \beta_{xq} \ln\left(\pi'_q\big|_{q=0} \hat{q}\right) + d$ and $\ln(\pi'_{\Delta x}\big|_{\Delta x=0} \widehat{\Delta x}) = \beta_{xq} \ln(\widehat{\pi_q}) + d$ respectively; where normalisation $\pi_q = 0$ at $q = 0$ and $\pi_{\Delta x} = 0$ at $\Delta x = 0$ is applied and is useful because these representations trivially arrives at the equilibrium defined by $\ln(\pi_{\Delta x}) = \beta_{xq} \ln(\pi_q) + d$ when $\hat{q} = \frac{\pi_q}{\pi'_q\big|_{q=0}}$ and $\widehat{\Delta x} = \frac{\pi_{\Delta x}}{\pi'_{\Delta x}\big|_{\Delta x=0}}$. Importantly, these relationships offer the convenient economic interpretation of supply and demand curves. That is, the supply of resources from nature to the firm is represented by $\ln(\widehat{\pi_{\Delta x}}) = \beta_{xq} \ln(\hat{q}) + \left(d + \beta_{xq} \ln\left(\pi'_q\big|_{q=0}\right)\right)$. But, given that at all points in time $r_x \pi_{\Delta x} = r_q \pi_q$ and $r_x = \beta_{xq} r_q$ then $\pi_q = \beta_{xq} \pi_{\Delta x}$ which implies the supply can also be written as $\ln(\widehat{\pi_q}) = \beta_{xq} \ln(\hat{q}) + \left(d + \beta_{xq} \ln\left(\pi'_q\big|_{q=0}\right) + \ln(\beta_{xq})\right)$; restricting $\beta_{xq} > 0$ to ensure real-valued $\ln(\beta_{xq})$. The demand can in turn be represented by $\ln(\widehat{\pi_q}) = \frac{(\ln(\pi'_{\Delta x}\big|_{\Delta x=0}) - d + e)}{\beta_{xq}} - \frac{\ln(\hat{q})}{\beta_{xq}}$; making the simplification that an increase in the rate of natural resource flowing from nature to the firm can equally be considered as a decrease in the rate of that natural resource flowing from the firm to nature i.e., formally $\overline{\left(\dot{\ln(\widehat{\Delta x})}\right)} = -\overline{\left(\dot{\ln(\hat{q})}\right)}$ which implies $\ln(\widehat{\Delta x}) = -\ln(\hat{q}) + e$.

Graphical considerations relating to slope, constants and intercepts of supply and demand equations are useful to investigate at this stage. Notice that the slope of supply curve is $\beta_{xq}$ and that the intercept $d + \beta_{xq} \ln\left(\pi'_q\big|_{q=0}\right) + \ln(\beta_{xq})$ is a function of $\beta_{xq}$. To simplify the intercept and noting that $d$ may also be a function of $\beta_{xq}$ it is convenient to consider first order estimate $d(\beta_{xq}^{(0)}) = m \beta_{xq}^{(0)} + n$. Rearranging the supply curve equation and taking the expectation over $\ln(\widehat{\pi_q})$ and $\ln(\hat{q})$ obtains $d(\beta_{xq}^{(0)}) = \overline{\ln(\widehat{\pi_q})} - \beta_{xq}^{(0)} \overline{\ln(\hat{q})} - \beta_{xq}^{(0)} \ln\left(\pi'_q\big|_{q=0}\right) - \ln(\beta_{xq}^{(0)})$ as a function of a single variable $\beta_{xq}^{(0)}$. Taking the derivative $d'(\beta_{xq}^{(0)})$ at $\beta_{xq}^{(0)} = 1$ arrives at $m = d'(1) = -\overline{\ln(\hat{q})} - \ln\left(\pi'_q\big|_{q=0}\right) - 1$. Thus, first order $d(\beta_{xq}^{(0)}) = \left(-\overline{\ln(\hat{q})} - \ln\left(\pi'_q\big|_{q=0}\right) - 1\right) \beta_{xq}^{(0)} + n$ and noting that $\overline{ln(\widehat{\pi_q})} - 1\overline{ln(\hat{q})} - 1ln\left(\pi'_q\big|_{q=0}\right) - ln(1) = d(1) = \left(-\overline{ln(\hat{q})} - ln\left(\pi'_q\big|_{q=0}\right) - 1\right)1 + n$, implies $n = 1 + \overline{ln(\widehat{\pi_q})}$. Thus, first order $d(\beta_{xq}{}^{(0)}) = \left(-\overline{ln(\hat{q})} - ln\left(\pi'_q\big|_{q=0}\right) - 1\right)\beta_{xq}{}^{(0)} + 1 + \overline{ln(\widehat{\pi_q})}$ holds and the supply curve becomes $ln(\widehat{\pi_q}) = \beta_{xq}ln(\hat{q}) + \left(\left(-\overline{ln(\hat{q})} - ln\left(\pi'_q\big|_{q=0}\right) - 1\right)\beta_{xq}{}^{(0)} + 1 + \overline{ln(\widehat{\pi_q})} + \beta_{xq}ln\left(\pi'_q\big|_{q=0}\right) + ln(\beta_{xq})\right)$ which simplifies to $ln(\widehat{\pi_q}) = \beta_{xq}ln(\hat{q}) + \overline{ln(\widehat{\pi_q})} - \overline{ln(\hat{q})}\beta_{xq} + \left(1 - \beta_{xq} + ln(\beta_{xq})\right)$ for $\beta_{xq}{}^{(0)} = \beta_{xq}$. Simplifying the intercept $f(\beta_{xq}) = 1 - \beta_{xq} + ln(\beta_{xq}) = ln(\beta_{xq}e^{1-\beta_{xq}})$ as a linear function of $ln(\beta_{xq})$ arrives at $f(\beta_{xq}) = ln(\beta_{xq})$ and therefore the supply curve becomes $ln(\widehat{\pi_q}) = \beta_{xq}ln(\hat{q}) + \overline{ln(\widehat{\pi_q})} - \overline{ln(\hat{q})}\beta_{xq} + ln(\beta_{xq})$.

The demand curve $ln(\hat{q}) = (ln(\pi'_{\Delta x}|_{\Delta x=0}) - d + e) - \beta_{xq}ln(\widehat{\pi_q})$, may be transformed in a similar way. Consider the linear estimate $d(\beta_{xq}{}^{(0)}) = j\beta_{xq}{}^{(0)} + k$ of $d(\beta_{xq}{}^{(0)}) = ln(\pi'_{\Delta x}|_{\Delta x=0}) + e - \beta_{xq}{}^{(0)}\overline{ln(\widehat{\pi_q})} - \overline{ln(\hat{q})}$. Then $j = d'(1) = -\overline{ln(\widehat{\pi_q})}$ and $k = d(1) = ln(\pi'_{\Delta x}|_{\Delta x=0}) + e - \overline{ln(\hat{q})}$ implies $d(\beta_{xq}{}^{(0)}) = -\beta_{xq}{}^{(0)}\overline{ln(\widehat{\pi_q})} + ln(\pi'_{\Delta x}|_{\Delta x=0}) + e - \overline{ln(\hat{q})}$. Therefore, the demand curve becomes $ln(\hat{q}) = \left(ln(\pi'_{\Delta x}|_{\Delta x=0}) - \left(-\beta_{xq}{}^{(0)}\overline{ln(\widehat{\pi_q})} + ln(\pi'_{\Delta x}|_{\Delta x=0}) + e - \overline{ln(\hat{q})}\right) + e\right) - \beta_{xq}ln(\widehat{\pi_q})$ which is simplified as $ln(\hat{q}) = \overline{ln(\hat{q})} + \beta_{xq}\overline{ln(\widehat{\pi_q})} - \beta_{xq}ln(\widehat{\pi_q})$ at $\beta_{xq}{}^{(0)} = \beta_{xq}$.

The supply and demand curve may be more simply expressed as $y = \beta_{xq}x + ln(\beta_{xq})$ and $x = -\beta_{xq}y$ where $y = ln(\widehat{\pi_q}) - \overline{ln(\widehat{\pi_q})}$ and $x = ln(\hat{q}) - \overline{ln(\hat{q})}$ denote deviations from mean values.

The supply curve and demand curve form a simultaneous system of equations that have the equilibrium solution at $y^{(e)} = \frac{ln(\beta_{xq})}{1+\beta_{xq}{}^2}$ and $x^{(e)} = -\frac{\beta_{xq}ln(\beta_{xq})}{1+\beta_{xq}{}^2}$; where $x^{(e)} = -\beta_{xq}y^{(e)}$ i.e., the locus of equilibrium points $(\hat{q}^{(e)}, \widehat{\pi_q}^{(e)})$ varies with $\beta_{xq}$. This shows that for $\beta_{xq} > 1$, the equilibrium price elasticity of equilibrium demand is $|\beta_{xq}| > 1$ (i.e., elastic) and also that $y = ln(\widehat{\pi_q}^{(e)}) > \overline{ln(\widehat{\pi_q})}$ and $ln(\hat{q}^{(e)}) < \overline{ln(\hat{q})}$ (i.e., the price is relatively higher than a measure of the middle and the quantity is relatively lower – a reflection of higher opportunity cost payments and lower quantity usage required for nature goods that have relatively large changes in nature good quantity with respect to changes in firm related price). Interestingly the elasticity of the supply curve is the inverse of the asset beta whereas the elasticity of the demand curve is the asset beta. Thus, an increasing asset beta makes supply more inelastic and demand more elastic. Furthermore, $\beta_{xq} = 1$ implies $y^{(e)} = 0$ and $x^{(e)} = 0$, indicating that the mean values $\overline{ln(\hat{q})}$ and $\overline{ln(\widehat{\pi_q})}$ are the market equilibrium; which makes economic sense given that the natural resource has the same rate of return as the industry incumbent firms at $\beta_{xq} = 1$ and therefore the variation of natural resource price and quantity contributes no additional opportunity costs and thus the mean values are the appropriate ones. In addition, $\int_0^\infty y^{(e)}d\beta_{xq} = \int_0^\infty \frac{ln(\beta_{xq})}{1+\beta_{xq}{}^2}d\beta_{xq} = 0$ also makes economics sense and suggest that the supply curve intercept $ln(\beta_{xq})$ is appropriate to use, given that in an economy where $x^{(e)} = 0$ and where a unit of a natural resource can be assigned to each and all of the continuum of firms with different $\beta_{xq} \in (0, \infty)$, the sum of the opportunity cost payments must be zero given that no opportunities are foregone.

In general, the supply curve denotes that as price $\widehat{\pi_q}$ increases then the willingness to sell natural resource increases; to effectively pay the marginal damage cost in selling the resource for processing by firms. The demand curve denotes that as the price $\widehat{\pi_q}$ decreases, the quantity demanded by firms increases; with the price being the same as the marginal cost of control (by not using the marginal unit of resource in firm production). The price $\widehat{\pi_q}$ has the usual interpretation as a payment for the unit of resource for forgoing spending on other opportunities i.e., it is an opportunity cost. Moreover, the supply and demand curve may be utilized to estimate welfare related consumer, producer and total surplus values with the usual caveats e.g., marginal utility of money. Of substantive interest is the total user cost, as the product of the equilibrium quantity $\hat{q}^{(e)}$ and price $\widehat{\pi_q}^{(e)}$, to estimate the opportunity foregone at a macroeconomic level.

The general form of the supply and demand curves considered above is $\left(ln(\widehat{\pi_q}) - \overline{ln(\widehat{\pi_q})}\right) = \beta_{xq}\left(ln(\hat{q}) - \overline{ln(\hat{q})}\right) + ln(\beta_{xq})$ and $\left(ln(\hat{q}) - \overline{ln(\hat{q})}\right) = -\beta_{xq}\left(ln(\widehat{\pi_q}) - \overline{ln(\widehat{\pi_q})}\right)$ respectively; which results in equilibrium point $\left(\overline{ln(\hat{q})} - \frac{\beta_{xq} ln(\beta_{xq})}{1+\beta_{xq}^2}, \overline{ln(\widehat{\pi_q})} + \frac{ln(\beta_{xq})}{1+\beta_{xq}^2}\right)$ such that $\left(ln(\hat{q}) - \overline{ln(\hat{q})}\right)^{(e)} = -\beta_{xq}\left(ln(\widehat{\pi_q}) - \overline{ln(\widehat{\pi_q})}\right)^{(e)}$. Recall that restricting $\beta_{xq} > 0$ ensured real-valued $ln(\beta_{xq})$. Utilising the inherent symmetry (which is simply a mapping between the neighbouring firm-nature worlds (perspectives), that are simply denoted by $(ln(\hat{q}), ln(\widehat{\pi_q}), \beta_{xq})$ and $(ln(\widehat{\Delta x}), ln(\widehat{\pi_{\Delta x}}), \beta_{qx})$ respectively), the derivation starting from a nature oriented perspective culminates in $\left(ln(\widehat{\Delta x}) - \overline{ln(\widehat{\Delta x})}\right)^{(e)} = -\beta_{qx}\left(ln(\widehat{\pi_{\Delta x}}) - \overline{ln(\widehat{\pi_{\Delta x}})}\right)^{(e)}$. Recalling that $\beta_{xq}\pi^{(N)}_{\Delta x} = \pi_q$ implies $ln(\beta_{xq}) + ln(\widehat{\pi_{\Delta x}}) = ln(\widehat{\pi_q})$, $\beta_{xq}\beta_{qx} = 1$ and $ln(\widehat{\Delta x}) = -ln(\hat{q}) + e$, this may be rewritten as $\left(ln(\hat{q}) - \overline{ln(\hat{q})}\right)^{(e)} = \frac{1}{\beta_{xq}}\left(ln(\widehat{\pi_q}) - \overline{ln(\widehat{\pi_q})}\right)^{(e)}$. It is then trivial to see that, a world in which $\left(ln(\hat{q}) - \overline{ln(\hat{q})}\right)^{(e)}$ and $\left(ln(\widehat{\pi_q}) - \overline{ln(\widehat{\pi_q})}\right)^{(e)}$ are positively correlated (for which $\left(ln(\hat{q}) - \overline{ln(\hat{q})}\right)^{(e)} = \frac{1}{\beta_{xq}}\left(ln(\widehat{\pi_q}) - \overline{ln(\widehat{\pi_q})}\right)^{(e)}$ applies) can be transformed (and is equivalent) to a world in which $\left(ln(\hat{q}) - \overline{ln(\hat{q})}\right)^{(e)}$ and $\left(ln(\widehat{\pi_q}) - \overline{ln(\widehat{\pi_q})}\right)^{(e)}$ are negatively correlated (for which $\left(ln(\hat{q}) - \overline{ln(\hat{q})}\right)^{(e)} = -\beta_{xq}\left(ln(\widehat{\pi_q}) - \overline{ln(\widehat{\pi_q})}\right)^{(e)}$ applies); by simply substituting $\beta_{xq}$ for $\frac{1}{\beta_{xq}}$. In short, the symmetry provides for the equivalence principle that a world in which $\left(ln(\hat{q}) - \overline{ln(\hat{q})}\right)^{(e)}$ and $\left(ln(\widehat{\pi_q}) - \overline{ln(\widehat{\pi_q})}\right)^{(e)}$ are positively correlated can be translated into a negatively correlated world (for which the above exposition of reasoning applies) by substituting the inverse of the natural asset beta.

The equivalence principle facilitates the interpretation of the coefficients in regression $E\left[\left(ln(\hat{q}) - \overline{ln(\hat{q})}\right)^{(e)} | \left(ln(\widehat{\pi_q}) - \overline{ln(\widehat{\pi_q})}\right)^{(e)}\right] = \alpha_{\widehat{\pi_q}}\left(ln(\widehat{\pi_q}) - \overline{ln(\widehat{\pi_q})}\right)^{(e)} + \alpha_0$. Specifically, for $\alpha_{\widehat{\pi_q}} < 0$, $\left(ln(\hat{q}) - \overline{ln(\hat{q})}\right)^{(e)} = -\beta_{xq}\left(ln(\widehat{\pi_q}) - \overline{ln(\widehat{\pi_q})}\right)^{(e)}$ applies such that $\alpha_{\widehat{\pi_q}} = \beta_{xq}$ and for $\alpha_{\widehat{\pi_q}} > 0$, $\left(ln(\hat{q}) - \overline{ln(\hat{q})}\right)^{(e)} = \frac{1}{\beta_{xq}}\left(ln(\widehat{\pi_q}) - \overline{ln(\widehat{\pi_q})}\right)^{(e)}$ applies such that $\alpha_{\widehat{\pi_q}} = \frac{1}{\beta_{xq}}$. Moreover, incorporating shocks into the above specification of supply and demand curves provides a means of modelling and estimating the observed-real-world equilibrium prices that result from shocked supply and demand curve $\left(ln(\widehat{\pi_q}) - \overline{ln(\widehat{\pi_q})}\right) = \beta_{xq}\left(ln(\hat{q}) - \overline{ln(\hat{q})}\right) + ln(\beta_{xq}) + \epsilon_S$ and $\left(ln(\hat{q}) - \overline{ln(\hat{q})}\right) = -\beta_{xq}\left(ln(\widehat{\pi_q}) - \overline{ln(\widehat{\pi_q})}\right) + \epsilon_D$ respectively; thus, equilibrium prices $x^{(e)} =$

$-\beta_{xq}y^{(e)} + \epsilon_{(e)}$ where $\epsilon_{(e)} = \frac{\epsilon_D(1+\beta_{xq})}{1+\beta_{xq}^2}$ result. This formulation has the utility of lending itself to empirical testing towards a validation of the theory in the following section.

In summary, the supply and demand curves are informed by the firm's asset beta and the natural asset beta that are also used to estimate the rate of return for nature to firm and firm to nature investments respectively. The returns allow the estimation of the natural asset beta $\beta_{xq}$ associated with the variation in the natural resource flow returns with respect to the firm returns (rather than the entire market in natural asset beta $\beta_{xm}$). The nature-firm natural asset beta $\beta_{xq}$ directly informs the profiles of the supply and demand curves. The equilibrium at which the supply and demand curves meet identifies the market exchange price and quantity for the natural resource that ensures constant consumption sustainability; given the fundamental alignment of the approach with the Grainger (2024) economy.

## 5 Testing the approach proof-of-concept

A test application of the above theory implied approach with real world data is performed in the following to facilitate a proof of concept. Given the need for caution in any first application, the origin of the real-world data is not reported at this stage; given the anticipated refinements to the approach preceding a more rigorous application necessitates acting in a prudential manner by not commenting at this early stage on the real-world industry. Of primary interest, irrespective of real-world industry, is whether a natural asset beta can be calculated at all and if so, whether the beta informs supply and demand curves that achieve an equilibrium to estimate the optimal quantity and price of the associated natural resource. It is entirely possible that a natural asset beta is not statistically significant upon analysing the data and that the equilibrium point is not able to be estimated, thus the interest in this proof-of-concept test of the approach.

The dataset contained aggregate gross yearly cashflows of an industry that extracted resources from nature. The industry, as part of its operation, also invested to grow those natural resources. The supply and demand relationships between the opportunity cost $\widehat{\pi_q}$ and the resource flow $q$, to infer the natural asset beta value $\beta_{xq}$, are of interest for the proof-of-concept exercise. Simply put, $\widehat{\pi_q}$ represents the covariation between firm free cash flow and the quantity of natural resource flow; $\widehat{\pi_q}$ is typically a monetary denoted price, but more generally it can be any flow of value (e.g., another resource) that is traded off with $q$. For the purposes of testing, the component of yearly gross value data for a particular industry (denoted by vector $\vec{v}$) that, ceteris paribus, covaried with the yearly natural resource flow data (denoted by vector $\vec{q}$) was used to estimate $\widehat{\pi_q}$. Formally, $\widehat{\pi_q} = \frac{\vec{v}\cdot\vec{q}}{\|\vec{v}\|\|\vec{q}\|}\vec{v} \oslash \vec{q}$ as the component of $\vec{v}$ in the direction of vector $\vec{q}$ undergoes Hadamard element-wise division to arrive at the value per (natural resource) unit dimension of $\widehat{\pi_q}$; a notion that has links to statistical geometry concepts of Bryant (1984).

Figure 1 depicts the linear regression estimate of the natural asset beta using mean centred covariates associated with $ln(\widehat{\pi_q})$ and $ln(\widehat{\Delta q})$ i.e., $y^{(e)}$ and $x^{(e)}$ respectively. Specifically, a control function approach, along the lines of Heckman and Navarro-Lozano (2004), was used to control for potential endogeneity and confounding by utilising four weakly exogenous instruments for $y^{(e)}$; with posthoc tests for RESET and residuals validating both an appropriate functional form and normality assumptions are satisfied. Notice that $\beta_{qx}$ has a statistically significant point estimate of 0.919. Because $\beta_{xm} = \beta_{xq}\beta_{qm}$, $\beta_{xq} = 0.919$ and that market estimates of $\beta_{qm} = 5.36$ (as the covariation of price returns for the resource investment operation with the returns of a market index) and $r_m = 2.9\%$ per year, then the natural asset beta is $\beta_{xm} = 4.93$ and its associated rate of return is $r_x = 14.3\%$ per year.

**Figure 1: Linear regression to estimate the natural asset beta**

| $x^{(e)} = \widehat{ln\Delta q} - \overline{ln(\widehat{\Delta q})}$ | Coef. | St.Err. | t-value | p-value | [95% Conf | Interval] | Sig |
|---|---|---|---|---|---|---|---|
| $y^{(e)} = ln(\widehat{\pi_q}) - \overline{ln(\widehat{\pi_q})}$ | -.919 | .018 | -50.36 | 0 | -.958 | -.881 | *** |
| Control fn | -.089 | .065 | -1.36 | .191 | -.227 | .049 | |
| Constant | 0 | .033 | 0.00 | 1 | -.069 | .069 | |
| | | | | | | | |
| Mean dependent var | | 0.000 | SD dependent var | | | 1.778 | |
| R-squared | | 0.994 | Number of obs | | | 19 | |
| F-test | | 1398.635 | Prob > F | | | 0.000 | |
| Akaike crit. (AIC) | | -17.454 | Bayesian crit. (BIC) | | | -14.621 | |

*** $p<.01$, ** $p<.05$, * $p<.1$

The equations of the supply and demand curves can thus be estimated as $\left(ln(\widehat{\pi_q}) - \overline{ln(\widehat{\pi_q})}\right) = \beta_{xq}\left(ln(\widehat{\Delta q}) - \overline{ln(\widehat{\Delta q})}\right) + ln(\beta_{xq})$ and $\left(ln(\widehat{\Delta q}) - \overline{ln(\widehat{\Delta q})}\right) = -\beta_{xq}\left(ln(\widehat{\pi_q}) - \overline{ln(\widehat{\pi_q})}\right)$ substituting for $\beta_{xq} = 0.919$ and Figure 2's $\overline{ln(\widehat{\Delta q})} = 2.113$ and $\overline{ln(\widehat{\pi_q})} = 2.828$ results in the estimation of the equilibrium price $\widehat{\pi_q}^{(e)} = e^{2.782}$ and quantity $\widehat{\Delta q}^{(e)} = e^{2.155}$; implying a total user cost of $\widehat{\pi_q}^{(e)}\widehat{\Delta q}^{(e)} = e^{2.782+2.155}$.

**Figure 2: Descriptive Statistics**

| Variable | Obs | Mean | Std. Dev. | Min | Max |
|---|---|---|---|---|---|
| $ln(\widehat{\Delta q})$ | 19 | 2.113 | 1.778 | -.357 | 4.461 |
| $ln(\widehat{\pi_q})$ | 19 | 2.828 | 1.913 | .361 | 5.585 |

The use of these natural resource by the firm, relative to keeping resources in-situ (in nature), is equivalent to taking on debt with a requirement to pay 14.3% per year. Alternatively, the 14.3% per year payment can be seen as equivalent to a perpetuity of insurance premium payments associated with insurance cover amounting to the present value of the natural resource used. Notably, the calculated equilibrium price and quantity are those at which constant consumption sustainability is maintained; being within the observed range of Figure 2. The equilibrium price is lower than the mean and the equilibrium quantity is higher than the mean. The firm effectively utilises a quantity of natural resources $ln(\widehat{\Delta q})$, in this particular case to expand the natural resource stock, but in so doing forgoes opportunities; hence the opportunity cost $ln(\widehat{\pi_q})$. More broadly in this particular case, the supply curve can be seen to reflect the opportunities foregone by this expansion activity and the demand curve can be seen to reflect the opportunities foregone by opposing it (i.e., acting instead to contract); which results in an equilibrium.

Noteworthy, is that while point estimates are sufficient for the expositive purposes of this paper, the interval of the equilibrium price $ln(\widehat{\pi_q})$ and quantity $ln(\widehat{\Delta q})$, total user cost $ln(\widehat{\pi_q \Delta q})$, natural asset beta with respect to the market $\beta_{xm}$ and the associated rate of return $r_x$ are useful in practice. The intervals can be generated, by assuming that the variation results from normally distributed $\beta_{xq}$, for which the 90 percent confidence interval may be numerically generated as depicted in Figure 3.

**Figure 3: 90% confidence interval of estimates**

| Between | $ln(\widehat{\pi_q})$ | $ln(\widehat{\Delta q})$ | $ln(\widehat{\pi_q \Delta q})$ | $\beta_{xm}$ | $r_x$ |
|---|---|---|---|---|---|
| Minimum | 2.76 | 2.14 | 4.93 | 4.80 | 13.9 |
| Maximum | 2.81 | 2.17 | 4.94 | 5.15 | 14.9 |

These estimates, for the given dataset, can provide useful quantitative details for decision-making, and the following example is intended to give an indication of the benefits brought. Consider a typical setting in which the operational cost of expanding the natural resource is known, but the market price an investor would pay for the expansion not known. If the operational cost of the natural resource expansion was greater than the market price an investor would pay, then the expansion should obviously not have occurred; hence the importance of estimating the market price. The market price estimate $\widehat{\pi_q}^{(e)}$ then provides the firm with an upper bound for the operation cost per unit of the expansion. It also provides the necessary price signals to inform investor buy or sell decisions; including how much they should allocate to the firm in light of the investment and their broader portfolio, whether the firm suffers perquisite consumption, or if arbitrage opportunities exist in the market.

In addition to guiding the allocation of finance across investments at a point in time, the estimated rate of return $r_x$ provides an informative measure of cost of delaying the use of the resource. Obviously, the natural asset beta $\beta_{xq}$ underpins both of these estimates and can be used to price more exotic (state of the world dependent) payouts, but it is sufficient here to show that it can at least estimate natural resource prices (previously in the shadows) to guide contemporaneous (across space) investment and effort allocations as well as intertemporal (across time) investment and effort allocations. Simply put, the natural asset beta completes the market by utilising the covariation of nature based physical flows and economy based monetary flows to estimate natural resource market exchange prices and rates of returns to inform decision-making by investors, firms and policymakers.

In short, the approach has, for this dataset, been able to estimate the natural asset beta, the associated natural asset return, the optimal equilibrium quantity and price to ensure constant consumption sustainability. Importantly, estimates of natural asset beta, interest rate, market price and optimal quantity provide information to inform trade-off decisions across space (i.e., trade-offs across current investment decisions) and time (i.e., intertemporal investment decision trade-offs). However, the use of point estimates needs to be expanded upon in future work to assess the range of values associated with these estimates, particularly for the equilibrium price. Despite this limitation, which also haunts the usual firm asset beta, one can imagine the utility of a database of natural asset betas to facilitate the valuation of natural resources in a manner similar to that seen in corporate finance valuations. The additional advantage of this approach is that it incorporates constant consumption sustainability notions at the very outset of the derivation of the natural asset beta.

## 6 Concluding remarks

The economic model developed in this paper shows that it is indeed possible to develop a corporate finance asset beta notion for natural resources. The endogenous discounting of Hartwick and Van Long (2020) and the rate of return link to market exchange prices of Grainger (2024) provides the pivotal foundation for considering the covariations of rates of returns. The subsequent proof of concept test of the approach developed in this paper shows promise. Specifically, it finds that a natural asset beta can be estimated and employed to find the associated rate or return, market exchange price and quantity that ensure constant consumption sustainability as originally defined in Solow (1974). Despite these novel contributions to the literature, the work proposes that future research would be well-placed to consider the sensitivities of the estimates calculated by using the proposed approach.

The reported research is aligned with the current appetite and momentum of global efforts to incorporate monetary values of natural resources into the system of national accounts (Hein et al., 2020). Importantly, it directly addresses the open problem of converting physical accounting of natural resources into monetary ones. As the Sharpe (1964) capital asset pricing model provided a means to operationalise Markowitz (1991) portfolio theory, the present work anticipates a similar operationalization of the underpinning theory of constant consumption sustainability; particularly given the alignment with the currently real-world dominant enterprise valuation practices remain substantively aligned with Sharpe (1964). Finally, it is hoped that the work facilitates uptake of the mass of research and endeavours relating to nature to inform trade-off decisions that ensure not too much or too little of nature is utilised by humans.